%% file: main.tex
\documentclass[conference]{IEEEtran}
\IEEEoverridecommandlockouts
\usepackage[utf8]{inputenc}
\usepackage{subfig}
\usepackage{cite}
\usepackage{amsmath,amssymb,amsfonts}
\usepackage{algorithmic}
\usepackage{graphicx}
\usepackage{textcomp}
\usepackage{xcolor}
\usepackage{physics}
\usepackage[capitalise]{cleveref} 

\def\BibTeX{{\rm B\kern-.05em{\sc i\kern-.025em b}\kern-.08em
    T\kern-.1667em\lower.7ex\hbox{E}\kern-.125emX}}
    
\crefname{section}{Sec.}{Secs.}

\newcommand{\ec}{E_C}
\newcommand{\ej}{E_J}
\newcommand{\el}{E_L}
\newcommand{\chn}{\hat{n}}
\newcommand{\fxn}{\hat{\phi}}
\newcommand{\bfield}{\varphi_e}
\newcommand{\ar}{\hat{a}}
\newcommand{\ad}{\hat{a}^\dagger}
\newcommand{\ham}{\hat{H}}
\newcommand{\tipangle}{\delta_\theta}
\newcommand{\tipanglez}{\delta_z}
\newcommand{\todo}[1]{{\color{red}{!!!}}}

\newcommand{\tgt}{T}
\newcommand{\subu}{\tilde{U}}

\newcommand{\iswap}{\text{iSWAP}}
\newcommand{\cz}{\text{CZ}}
\def\rz/{\emph{fid2}}
\def\avg/{\emph{fid1}}
\def\rzf/{\emph{Z-independent gate fidelity}}
\def\avgf/{\emph{subspace-averaged gate fidelity}}

\makeatletter
\def\ps@IEEEtitlepagestyle{%
  \def\@oddfoot{\mycopyrightnotice}%
}
\def\mycopyrightnotice{%
  \begin{minipage}{\textwidth}
  \centering \scriptsize
  Copyright~\copyright~2021 IEEE. Personal use of this material is permitted. Permission from IEEE must be obtained for all other uses, in any current or future media, including reprinting/republishing this material for advertising or promotional purposes, creating new collective works, for resale or redistribution to servers or lists, or reuse of any copyrighted component of this work in other works.
  \end{minipage}
}
\makeatother

\begin{document}

\title{Practical implications of SFQ-based two-qubit gates\\
\thanks{This work is funded in part by EPiQC, an NSF Expedition in Computing, under grants CCF-1730082/1730449; in part by STAQ under grant NSF Phy-1818914; in part by DOE grants DE-SC0020289 and DE-SC0020331; and in part by NSF OMA-2016136 and the Q-NEXT DOE NQI Center.}
}

\author{\IEEEauthorblockN{ Mohammad Reza Jokar\textsuperscript{*} \thanks{\textsuperscript{*}These two authors contributed equally.}}
\IEEEauthorblockA{\textit{Computer Science Department} \\
\textit{University of Chicago}\\
Chicago, USA \\
jokar@uchicago.edu}
\and
\IEEEauthorblockN{Richard Rines\textsuperscript{*}}
\IEEEauthorblockA{\textit{Computer Science Department} \\
\textit{University of Chicago}\\
Chicago, USA \\
richrines@uchicago.edu}
\and
\IEEEauthorblockN{Frederic T. Chong\textsuperscript{†} \thanks{\textsuperscript{†}Disclosure:  Fred Chong is Chief Scientist at Super.tech and an advisor to Quantum Circuits, Inc.}}
\IEEEauthorblockA{\textit{Computer Science Department} \\
\textit{University of Chicago}\\
Chicago, USA \\
chong@cs.uchicago.edu}
\and
}

\maketitle

\begin{abstract}
Scalability of today's superconducting quantum computers is limited due to the huge costs of generating/routing microwave control pulses per qubit from room temperature. One active research area in both industry and academia is to push the classical controllers to the dilution refrigerator in order to increase the scalability of quantum computers. Superconducting Single Flux Quantum (SFQ) is a classical logic technology with low power consumption and ultra-high speed, and thus is a promising candidate for in-fridge classical controllers with maximized scalability. Prior work has demonstrated high-fidelity SFQ-based single-qubit gates. However, little research has been done on SFQ-based multi-qubit gates, which are necessary to realize SFQ-based universal quantum computing.

In this paper, we present the first thorough analysis of SFQ-based two-qubit gates. Our observations show that SFQ-based two-qubit gates tend to have high leakage to qubit non-computational subspace, which presents severe design challenges. We show that despite these challenges, we can realize gates with high fidelity by carefully designing optimal control methods and qubit architectures. We develop optimal control methods that suppress leakage, and also investigate various qubit architectures that reduce the leakage. After carefully engineering our SFQ-friendly quantum system, we show that it can achieve similar gate fidelity and gate time to microwave-based quantum systems. The promising results of this paper show that (1) SFQ-based universal quantum computation is both feasible and effective; and (2) SFQ is a promising approach in designing classical controller for quantum machines because it can increase the scalability while preserving gate fidelity and performance.
\end{abstract}


\section{Introduction}
\label{sec:intro}
\input{tex/intro.tex}
\section{Background and Motivation}
\label{sec:back}
\input{tex/background.tex}
\section{Detailed study of SFQ-based two-qubit gates}
\label{sec:study}
\input{tex/study.tex}
\section{Conclusion}
\label{sec:conclusion}
\input{tex/conclusion}

\section*{Acknowledgment}

We gratefully thank Kangbo Li (University of Wisconsin-Madison), Mogens Dalgaard (Aarhus University), and Jacob Sherson (Aarhus University) for valuable discussions and sharing their codes with us. We also thank Pranav Gokhale (Super.tech) for useful discussions. This work was completed in part with resources provided by the University of Chicago’s Research Computing Center.

\bibliographystyle{IEEEtranS}
\bibliography{references}

\end{document}

%% file: tex/intro.tex
A great milestone in quantum computing is the recent development of quantum computer prototypes thanks to great efforts in industry and academia. Superconducting quantum computing is one of the most promising technologies to realize a quantum computer, having been used to realize prototypes with $<$100 qubits \cite{ibm_device,ibm_machine,qarch_fu,qarch_yuan,google_machine1,google_machine2}. These prototypes rely on sending analog microwave signals per qubit from a classical controller at room temperature to the quantum chip inside a dilution refrigerator in order to perform quantum operations. Unfortunately, this scheme introduces severe scalability challenges due to high costs of electronics that are used to generate the microwave signals at room temperature, as well as heat dissipation inside the dilution refrigerator caused by routing the high-bandwidth signals to the quantum chip \cite{mcdermott_full,mcdermott_fab,mcdermott_kangbo}. Thus, design decisions must be made to address the scalability challenges of today's quantum computer prototypes and realize large-scale quantum computers, which are essential in running many quantum algorithms and performing quantum error correction.

One active research area in industry and academia is designing in-fridge classical controllers, which increase the scalability of quantum machines by generating and routing the control signals locally. Due to maturity of CMOS logic, Cryo-CMOS is one attractive logic technology to build in-fridge controllers. Prior work has demonstrated Cryo-CMOS controller prototypes that generate microwave control pulses inside the dilution refrigerator, and can scale to hundreds of qubits given the power budget of dilution refrigerators \cite{horseridge2}. 
Meanwhile, Superconducting Single Flux Quantum (SFQ) is proposed as an alternative logic technology in the literature. SFQ logic is less mature than CMOS but can maximize the scalability of in-fridge controllers due to its very low power consumption and ultra-high speed \cite{mcdermott_fab,mcdermott_full,mcdermott_kangbo,sfq_genetic}.

SFQ-based controllers can perform quantum operations by generating a train of SFQ pulses (instead of microwave control waveforms) inside the dilution refrigerator and applying them directly to the qubits \cite{mcdermott_kangbo,sfq_genetic}. Previous work has demonstrated high-fidelity single-qubit gates with low leakage to the non-computational subspace using SFQ pulses \cite{mcdermott_fab,mcdermott_kangbo}. Prior work also demonstrated SQF-based two-qubit gates considering a model which takes into account only the first two energy levels of the qubits (i.e., qubit computational subspace) \cite{alphazero}. However, there is a lack of a detailed analysis in the literature on high-fidelity SFQ-based two-qubit gates which takes into account leakage to the non-computational subspace. A key unanswered question is: are SFQ-based two-qubit gates with high fidelity and low leakage feasible and effective? In this paper, we present the first thorough study on SFQ-based two-qubit gates, and demonstrate that we can realize them with high fidelity and low leakage by carefully designing our quantum optimal control method and qubit architecture.

We first demonstrate that it is essential to take higher energy levels of the qubits into consideration in our optimal control method. Similar study has been done in the literature on SFQ-based single-qubit gates \cite{sfq_genetic} where the authors show that taking into account the first three lowest energy levels (i.e., the qubit computational subspace and one higher energy level) in the optimal control method is sufficient to find high-fidelity gates with low leakage to higher energy levels. In this paper, we show that two-qubit gates have much higher tendency to leak to higher energy levels, thus it is challenging to find high-fidelity gates even if we take into account up to five energy levels in our optimal control method. Thus, we must take further steps by developing SFQ-based optimal control methods to suppress leakage and investigating qubit architectures and configurations that reduce leakage.

We first study transmon qubit devices with $\Omega_{x}$ control fields, which are widely used in both SFQ-based and microwave-based systems \cite{mcdermott_kangbo,alphazero,sgrape}. We show that we can realize high-fidelity SFQ-based two-qubit gates with low leakage to higher energy levels. 
We then investigate two possible extensions of this design in order to reduce the gate time while keeping the leakage low: (1) the addition of $\sigma_z$ SFQ control pulses implemented via frequency-tunable split-transmon devices; and (2) the use of SFQ control pulses in combination with high-anharmonicity fluxonium qubits.

Finally, we compare our SFQ-friendly quantum system with microwave-based quantum systems, and show that we can achieve similar gate fidelity and gate time using SFQ. This shows that SFQ is a promising approach to implement classical controllers for quantum computers because it can maximize the scalability of quantum computers due to the unique characteristics of SFQ logic, while delivering similar fidelity and performance to that of state-of-the-art microwave-based systems. 

To summarize, our key contributions are as follows:

\begin{itemize}
    \item We present the first study of SFQ-based two-qubit gates that takes into consideration the leakage to higher energy levels.
    \item We identify and discuss the main challenge in realizing high-fidelity SFQ-based two-qubit gates, which is high leakage to non-computational qubit subspace.
    \item We develop optimal control methods that suppress the population of higher energy levels in two-qubit gates.
    \item We study various qubit architectures and configurations in an attempt to engineer a quantum system with low leakage.
    \item We engineer an SFQ-friendly quantum system, and show that it can achieve similar gate fidelity as that of microwave-based system -- a promising result.
\end{itemize}

The rest of the paper is organized as follows. Sec. \ref{sec:back} presents a background on qubit architectures and configurations, quantum optimal control and SFQ-based gates, followed by a discussion on the motivation of our paper. Sec. \ref{sec:study} presents our methodology and the results of our detailed study on SFQ-based two-qubit gates, followed by a comparison with microwave-based two-qubit gates. Finally, Sec. \ref{sec:conclusion} concludes the paper.

%% file: tex/background.tex
Here we provide details of the physical systems we are targeting in order to distill the basic toolbox of quantum operations available to us for implementing high-performance SFQ-based quantum gates.  We motivate our analysis by describing the challenges of implementing high-fidelity gates on realistic quantum systems, the existing strategies for overcoming them on systems with analog control, and prior work on SFQ-based gates aiming to do the same.

\subsection{Physical system}
\label{sec:bg:system}

The evolution of a quantum system is governed by its Hamiltonian. For universal quantum computation, we require that the system provide (1) well-defined \emph{qubits}, or separable two-level quantum subsystems which can be independently initialized and measured; (2) a mechanism for generating entanglement between these qubits; and (3) a method for precisely controlling the system's evolution \cite{DiVincenzo2000}.
For the purposes of this investigation, we consider pairs of statically-coupled superconducting qubits, with the overall system Hamiltonian,
\begin{equation}
\label{eq:bg:hsum}
\ham = \sum_q \ham_{q} + \sum_q\ham_{q,d}(t) + \ham_{qq},
\end{equation}
where $\ham_q$ are the static Hamiltonian of each qubit, $\ham_{q,d}$ are the contributions of the time-dependent control signals applied to each qubit, and $\ham_{qq}$ is contributed by the inter-qubit coupling (and therefore is responsible for entanglement generation).
In the following, we express each of these terms for various hardware configurations in terms of the conjugate flux and charge number quantum operators $\fxn$ and $\chn$, where $[\chn,\fxn]=i$. 

\subsubsection{Transmons}

The superconducting transmon qubit comprises a Josephson junction (JJ) shunted to ground with a capacitor in order to minimize its sensitivity to charge noise \cite{transmons1,transmons2}.
The transmon Hamiltonian can be written as \cite{quantum_eng},
\begin{equation}
\label{eq:bg:tmon}
\ham_q = 4\ec \chn^2 - \ej \cos\fxn,
\end{equation}
where $\ec=e^2/2C_q$ indicates the capacitive energy (with $C_q$ including both the shunt capacitance and that of the JJ), $\ej=I_c\Phi_0/2\pi$ is the Josephson energy of a transmon with critical current $I_c$, and $\Phi_0$ is the magnetic flux quantum.

The spectrum of the single-transmon system can be found by diagonalizing \cref{eq:bg:tmon}.
For $\ec\ll\ej$ the transmon Hamiltonian can be expanded in the SHO Fock state basis,
\begin{equation}
\ham_q \approx \omega_{01} \ad\ar + \frac{\alpha}{2}\ad\ar(\ad\ar-1),
\end{equation}
where $\omega_{01}=\sqrt{8\ej\ec}-\ec$ is the qubit's oscillation frequency (that is, the energy gap between the ground and first excited state), $\alpha=\omega_{12}-\omega_{01}=-\ec$ is its anharmonicity, and we have made the substitution,
\begin{equation}
 \chn=\sqrt[4]{\ej/32\ec}\qty(\ar+\ad),
  \quad
 \fxn=i\sqrt[4]{2\ec/\ej}\qty(\ar-\ad),
\end{equation}
using the standard creation (annihilation) operators $\ad$ ($\ar$).
Typical transmon qubits are configured with oscillation frequencies $\omega_{01}/2\pi$ between 3 and 6 GHz and anharmonicity $\alpha/2\pi$ between 100 and 300 MHz \cite{quantum_eng}.
The nonzero anharmonicity makes it possible to isolate and address the system's $\{\ket0,\ket1\}$ subspace, providing the required well-defined two-level qubit.

\subsubsection{Frequency-tunable transmons}

The single-JJ transmon's oscillation frequency is fixed by its hardware components. We can instead construct a \emph{frequency-tunable} transmon by splitting its single JJ into a pair of parallel junctions (dc-SQUID) and driving an external magnetic flux $\bfield$ through the enclosed loop.
In this case the junction energy $\ej$ in \cref{eq:bg:tmon} is replaced with the flux-dependent effective energy \cite{quantum_eng},
\begin{equation}
  \ej' = \sqrt{(E_{J1}+E_{J2})^2\cos^2\bfield+\abs{E_{J1}-E_{J2}}^2\sin^2\bfield},
\end{equation}
where $E_{J1,2}$ are the Josephson energies of the respective JJs and $\bfield$ is the applied flux in units of $\Phi_0/\pi$.
The applied flux can then be used to tune the qubit's oscillation frequency, or equivalently implement $z$-axis rotations of the qubit.

For multi-qubit systems, flux control has also been employed to implement two-qubit gates by inducing resonant oscillations between multi-qubit states. For example, by bringing the qubit frequencies together, coherent oscillations between the $\ket{01}$ and $\ket{10}$ state will generate the $\iswap$ (or $\sqrt\iswap$) gate, whereas a $\cz$ gate can be implemented using the resonance between the $\ket{11}$ and $\ket{02}$ (or $\ket{20}$) states.
The latter case takes advantage of the higher energy levels of the transmon system, allowing the quantum state to temporarily leave the two-level qubit subspace during the execution of the gate.
Frequency-tunable transmons enable fast resonant two-qubit operations while decreasing crosstalk by allowing noninteracting qubits to be ``parked'' at well-separated oscillation frequencies.
This tunability comes at the cost of added complexity and sensitivity to magnetic flux noise.

\subsubsection{Fluxonium}

Though the transmon's nonzero anharmonicity makes it possible to target the two-level (qubit) subspace for quantum computation, its weakness relative to the oscillation frequency makes it prone to leakage to higher level states.
Alternative qubit technologies such as fluxonium \cite{Manucharyan2009} have been shown to increase anharmonicity with minimal cost in terms of noise sensitivity. The fluxonium qubit is constructed similarly to the transmon, but with an additional inductive shunt to ground implemented using an array of Josephson junctions connected in series.
The resulting Hamiltonian is written as \cite{quantum_eng},
\begin{equation}
\label{eq:bg:fluxonium}
\ham_q = 4\ec\chn^2 + \el\fxn^2 - \ej\cos(\fxn+\bfield),
\end{equation}
where $\el\ll\ej$ is the inductive energy of the junction array and $\bfield$ is an external magnetic flux through the qubit loop.

Fluxonium's sensitivity to flux noise is minimized at $\bfield=0$ and $\bfield=\pi$, where symmetry ensures that the energy dependence on $\bfield$ vanishes to first order. In the latter case, the qubit's oscillation frequency $\omega_{01}$ is significantly reduced relative to that of the subsequent transition ($\omega_{12}$), resulting in large, positive anharmonicity.
It is less trivial to approximate the fluxonium spectrum analytically; instead we diagonalize \cref{eq:bg:fluxonium} numerically to determine the computational basis states and energy spectrum of our system.
With typical hardware configurations, fluxonium qubits at $\bfield=\pi$ have $\omega_{01}\sim1$ GHz, while $\omega_{12}$ is 2-5 times larger.
For remainder of this paper, we assume that fluxonium is operating with a $\bfield=\pi$ static bias flux.

\subsubsection{Coupling}

We focus on systems with static coupling between qubits, such that the interaction Hamiltonian $H_{qq}$ is constant and uncontrollable (as opposed to, for example, tunable coupling systems \cite{google_machine1} which allow the interaction to be switched on and off on-demand but which would complicate the implementation of an SFQ-based controller). For superconducting qubits coupled via a capacitance $C_{qq}$, the coupling Hamiltonian in \cref{eq:bg:hsum} is,
\begin{equation}
  \label{eq:bg:coupler}
  \ham_{qq} = g_{qq} \chn_{q_0}\chn_{q_1},
\end{equation}
where $g_{qq} = 4e^2{C_{qq}}/{C_{q_0}C_{q_1}}$ quantifies the coupling strength.
Expressing \cref{eq:bg:hdrive} in the energy-basis rest frame of the undriven qubit, the dominant matrix elements of the coupling Hamiltonian (after the rotating wave approximation) are,
\begin{align}
\label{eq:bg:rfcoupler}
\ham_{qq}^{rf}(t) = J\sum_k & c^{(0)}_{k-1,k}c^{(1)}_{l,l-1} 
e^{i(\omega_{k,k-1}^{(0)}-\omega^{(1)}_{l-1,l})t} 
\nonumber
\\
\times\; & \big( \ketbra{k,l-1}{k-1,l} + h.c..\big),
\end{align}
where $J$ is a normalized coupling constant, and $c_{k,k-1}=c_{k-1,k}\approx\sqrt{k}/2$ for transmons whereas for fluxonium can be computed numerically by diagonalizing each qubit's Hamiltonian (\cref{eq:bg:fluxonium}).
Though this interaction cannot be disabled, the \emph{effective} coupling strength
between qubits is inversely proportional to the separation between their oscillation frequencies due to destructive interference caused by time-averaging the rotating phase in \cref{eq:bg:rfcoupler}.
We can therefore preserve the independence of the qubits by designing the system such that the parking frequencies of coupled qubits are well separated.

\subsubsection{Drive}
\label{sec:bg:control}

The most common architecture for manipulating statically-coupled qubits is to apply microwave control signals directly to the qubits via a coupling capacitor. 
Given a time-dependent voltage source $V_d(t)$, the microwave drive Hamiltonian is,
\begin{equation}
\label{eq:bg:hdrive}
\ham_{q,d} = V_d(t) \frac{2eC_d}{C_d+C_q}\chn,
\end{equation}
where $C_d$ is the capacitance of the coupling capacitor.
Expressed in the rest frame of the qubit
and assuming a microwave drive $V_d(t)=\Omega_x(t)V_0\cos\omega_dt$ (where $\Omega_x(t)$ is the normalized pulse envelope and $V_0$ 
absorbs the details of the qubit and drive hardware),
\begin{equation}
\label{eq:bg:rfdrive} 
H_d^{rf}(t) = \Omega_x(t)\sum_kc_{k+1,k}e^{i(\omega_{k,k+1}-\omega_d)t}\ketbra{k+1}{k} + h.c..
\end{equation}
The time-dependent phases in \cref{eq:bg:rfdrive} allow us to selectively drive a given transition while others are suppressed by the time-dependent phase. For example, continuously driving with $\omega_d=\omega_{01}$ will drive Rabi oscillations in the qubit subspace while the qubit's nonzero anharmonicity $\omega_{12}-\omega_{01}=\alpha$ will suppress the $\ket1\leftrightarrow\ket2$ transition.
However, in order to have a finite gate time, the envelope $\Omega_x(t)$ must itself contain Fourier components which can diminish this suppression by overlapping with higher-order transitions, especially given the transmon's relatively small anharmonicity. Analytical pulse shaping models such as the DRAG scheme \cite{dragpulse} are therefore employed on microwave systems to precisely minimize the overlapping frequency components in the pulse shape.

Using the cross-resonance interaction \cite{crgate1,crgate2} it is also possible to induce two-qubit entangling operations with precisely detuned control signals applied to one or both qubits, making fixed-frequency transmons and microwave control sufficient for universal quantum computation. Successful quantum computer prototypes have been developed using this control mechanism \cite{ibm_device}.
On these systems, the speed of cross-resonance gates is proportional to the effective coupling between the qubits, creating a tradeoff between gate time and crosstalk.

\subsection{Microwave optimal control}

In practice, the broad control schemes outlined above are insufficient for high-precision quantum gates.
Analytical leakage suppression schemes are especially challenging for multi-qubit systems due to the exponentially increasing complexity of the energy spectrum and the contributions of each coupler.
This complexity is especially prevalent for cross-resonance gates, in which one transition between multi-qubit states is intentionally driven while all others must be suppressed.
Further, it is often desirable to allow the system to evolve outside the two-level subspace during the execution of the gate, as it provides more possible paths for realizing complicated operations within a short gate time
(as in the frequency-tunable CZ implementation described above).
Typical microwave systems therefore employ search-based optimal-control strategies such as the ubiquitous gradient ascent pulse engineering (GRAPE) tool \cite{sgrape} to generate pulse waveforms which implement quantum gates with high fidelity and low leakage.

\subsection{Fidelity functions}
\label{sec:fidelity}

Throughout this work, we quantify the performance of learned gates using two variants of average gate fidelity. In its general form, the average gate fidelity of a quantum operation $\mathcal{E}$ relative to a unitary target gate $\tgt$ is defined,
\begin{equation}
\label{eq:fidelity}
\overline{F}(\mathcal{E},T)
= \int d\psi \ev{\tgt^\dagger \mathcal{E}(\psi) \tgt}{\psi} ,
\end{equation}
where the average is over the normalized Haar distribution of quantum states.
Because we are interested only in how the gate affects \emph{qubits}, we 
would like our fidelity metric to (1) be agnostic to the behavior of the gate when applied to states outside the qubit subspace, and
(2) penalize gate-induced leakage from within the computational subspace.
We therefore define,
\begin{equation}
\mathcal{E}(\psi)  = \Pi U\Pi \ketbra\psi \Pi U^\dagger\Pi,
\end{equation}
where $U$ is the simulated (unitary) evolution including higher level states, and,
\begin{equation}
\Pi  = (\ketbra0 + \ketbra1)^{\otimes n},
\end{equation}
projects it into the qubit subspace.
The average in \cref{eq:fidelity} is then taken over just states $\ket\psi$ in the two-level subspace $SU(2^n)$, so that a \avgf/ can be calculated \cite{Ghosh2011},
\begin{equation}
\label{eq:2fidelity}
\overline{F_1}(U,T)
= \frac{ \tr\big(\Pi U\Pi U^\dagger\Pi\big) + \tr\big(\tgt\Pi U^\dagger\Pi\big)^2 }{2^{2n}+2^n},
\end{equation}

Because of the constrained control set available with SFQ control, we would further like to broaden our search target as much as possible. In practice, single-qubit $Z$ rotations can often be commuted through subsequent gates or implemented virtually. We therefore define the \rzf/, which is independent of trailing $Z$-rotations:
\begin{align}
\label{eq:rzfidelity}
\overline{F_2}(U,T)
&= \sup_{\vec\alpha} \overline{F_1}\big(Z_{\vec\alpha}\subu,\tgt\big),
\\
Z_{\vec\alpha}
&= Z(\alpha_1)\otimes\cdots\otimes Z(\alpha_n).
\end{align}

Finally, we can explicitly quantify leakage by computing the probability of measuring a state outside the qubit subspace after applying the gate to a state initially within that space. Averaging over the uniform distribution of all possible two-level input states, the \emph{average leakage} can be calculated,
\begin{equation}
\label{eq:leakage}
\overline{L}(U) 
= 1 - \frac{\tr(\Pi U \Pi U^\dagger \Pi)}{2^n}.
\end{equation}
From \cref{eq:leakage,eq:2fidelity,eq:rzfidelity} one can show that $\overline{F_1}(U,\tgt) 
\le \overline{F_2}(U,\tgt) 
\le 1-\overline{L}(U)$, so that as desired our fidelity metrics are upper-bound by the degree of leakage.

\begin{figure}[!t]
\begin{center}
\subfloat    {
\includegraphics[width=0.95\linewidth]{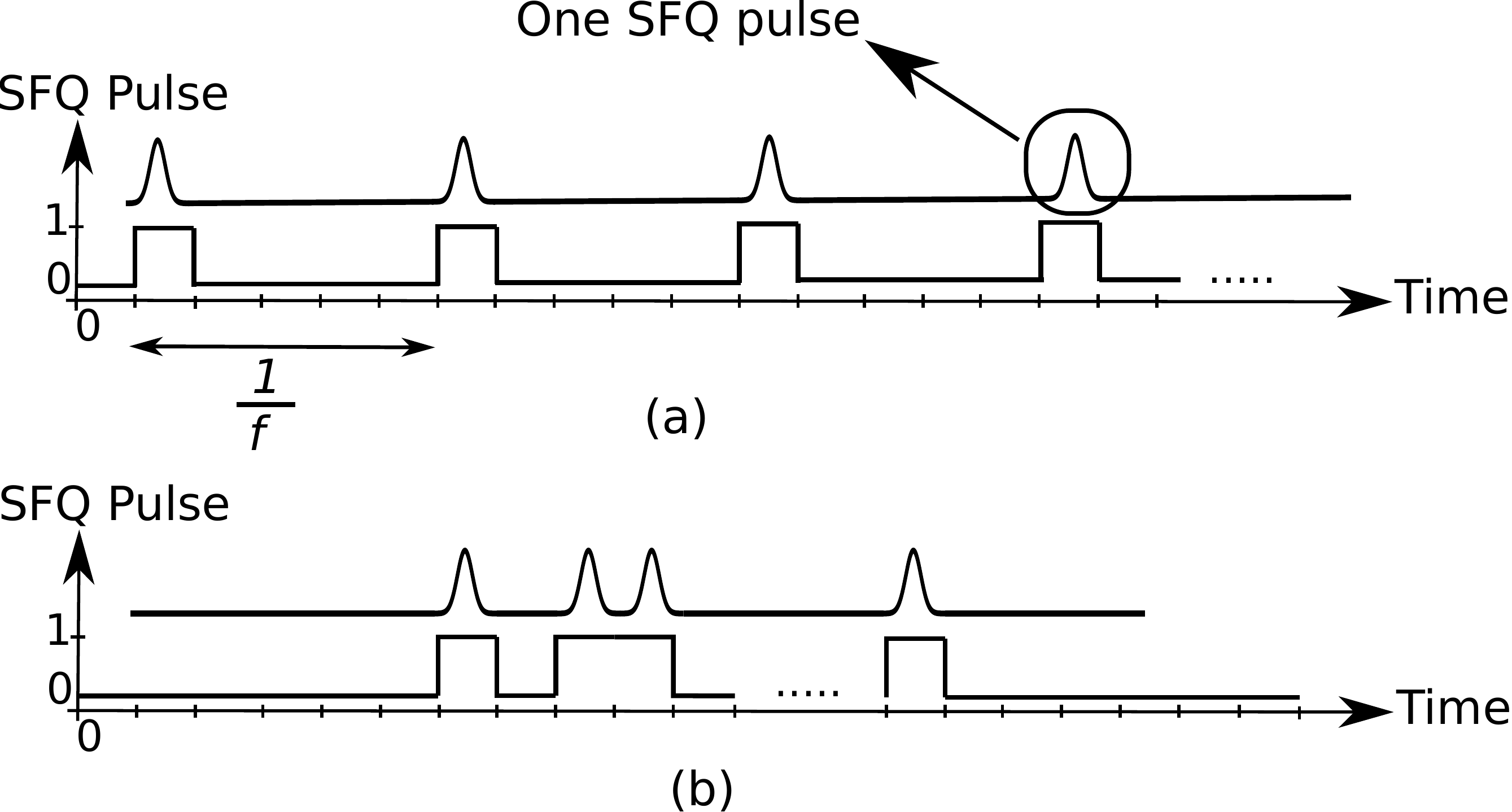}
}
\caption{Bit representation of SFQ pulse trains. (a) coherent pulses are applied to the qubit (1 pulse per qubit oscillation period) to perform rotations around the y axis. (b) a bitstream found by genetic algorithm to perform arbitrary unitary. Bitstreams are processed one bit at a time; if the bit is ``0'', no pulse is applied to the qubit, and if the bit is ``1'', one SFQ pulse is applied to the qubit.}
\label{fig:sfq}
\end{center}
\end{figure}

\captionsetup[subfigure]{labelformat=empty}
\begin{figure}[!t]
\begin{center}
\subfloat    {
\includegraphics[width=0.95\linewidth]{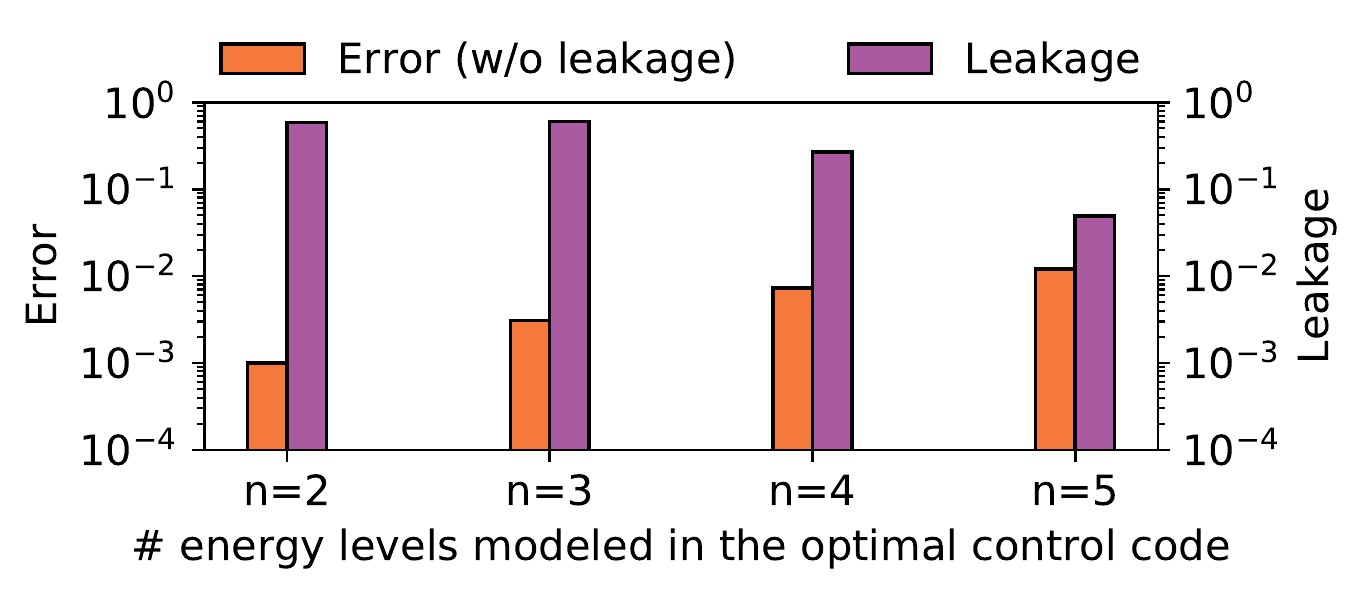}
}
\caption{Error and leakage of the best SFQ-based CZ gate with 20 ns gate time found by the genetic algorithm on transmon qubit devices with $\Omega_{x}$ control fields.
Error is computed using \cref{eq:2fidelity} and takes into account only the $n$ levels on which the gate was learned. Leakage is computed using \cref{eq:leakage} and considers higher levels.
Thus, low error does not necessarily translate to low leakage.}
\label{fig:mot}
\end{center}
\end{figure}

\subsection{SFQ control}

It has been proposed that quantum gates be implemented by applying SFQ pulses to the qubit directly in place of analog microwave control signals. 
The gate implementation is then described by a binary pulse train as shown in \cref{fig:sfq}, where in each cycle of the SFQ clock a pulse is either applied or not applied to each qubit. For two-qubit gates, we can apply different pulse trains to each qubit.

A single SFQ pulse is a rapid Gaussian voltage waveform,
\begin{equation}
V_d(t)=\tfrac{\Phi_0}{\sqrt{2\pi\tau^2}}e^{-t^2/2\tau^2},
\end{equation}
with a total area of exactly $\int dtV_d(t)=\Phi_0$ and a typical pulse width of $\tau=0.25$ ps.
Approximating $V_d(t)\approx\Phi_0\delta(t)$ and considering \cref{eq:bg:hdrive} in the energy-basis rest frame of the transmon, we expect a single pulse at time $t_0$ to implement the instantaneous gate,
\begin{equation}
\label{eq:bg:uflux}
U_{x}^{rf} = \exp \bigg\{ -i\tipangle \sum_k  e^{i\omega_{k,k+1}t_0}\sqrt{k}\ketbra{k}{k-1} + h.c.. \bigg\},
\end{equation}
where $\tipangle$ is the \emph{tip angle}, indicating the rotation angle induced by a single pulse in the qubit subspace. The tip angle is typically in the range of $10^{-3}$ to $10^{-1}$ radians \cite{mcdermott_kangbo,alphazero}, and is directly configurable via choices of qubit and coupling hardware; in our analysis we find that this configuration is extremely important for achieving high-performance SFQ gates.

In order to expand our narrow SFQ control toolset, we also consider an SFQ-based $\sigma_z$ operation for frequency-tunable transmons.
In this case, rather applying pulses to the qubit via a capacitive coupler, we assume that they are inductively coupled to the split transmon's dc-SQUID loop.
Approximating a single pulse as a delta function, the resulting gate is then simply,
\begin{equation}
  U_z = \sum_k e^{ik\tipanglez} \ketbra{k},
\end{equation}
where $\tipanglez$ is the $z$-axis tip angle determined by the hardware configuration.
In this case, achieving a non-negligible tip angle may require additional filter hardware in order to
both broaden the SFQ pulse shape and mitigate distortion caused by the mutual inductance between the qubit and the control line (as discussed in \cite{transmons2}).
With a rough calculation we find that $\tipanglez\sim0.03$ should easily be achievable using existing techniques (such as \cite{sfqfiltercircuits}).
Though on a single-qubit system $\sigma_z$ control would not be sufficient for universal quantum control, it turns out to be remarkably effective for realizing two-qubit gates when combined with the free evolution due to the static coupler.

Unlike the microwave drive, with SFQ pulses we cannot simply select a drive frequency in order to selectively drive a given transition while off-resonant transitions are suppressed by the rotating phase in \cref{eq:bg:uflux}. Instead, we are limited to selecting discrete clock cycles $t$ in which to apply $U_x(t)$.
If we constrain our system to the qubit subspace ($k\in\{0,1\}$), \cref{eq:bg:uflux} is simply a unitary rotation $e^{-i\tipangle(\cos(\omega t) X+\sin(\omega t) Y)}$ by angle $\tipangle$ about a time-dependent axis on the xy-plane.  
In this case, the problem is reduced to one of single-qubit gate composition (taking as basis gates the set of lab-frame single-clock-cycle unitaries generated by applying pulses to each possible subset of qubits), for which many analytic and search methods have been studied.
Empirically, in both prior work and our own examination it appears that pulse trains implementing high-fidelity, low leakage single-qubit gates are still readily discoverable when we model the system with additional energy states \cite{mcdermott_kangbo,mcdermott_full,sfq_genetic}.
This is perhaps unsurprising observing \cref{eq:bg:uflux}; though each pulse may result in some population transfer out of the qubit subspace, the simple energy spectrum of a single qubit near its ground state
makes it reasonable to expect symmetries to exist in which pairs or small groups of pulses will generate destructive interference in the non-qubit subspace (in fact, such symmetries were employed explicitly as part of the search algorithm outlined in \cite{mcdermott_kangbo}).

\subsection{Prior work on SFQ-based gates and the motivation of this paper}
\label{sec:motivation}

There has been detailed analysis of SFQ-based single-qubit gates in the literature \cite{mcdermott_full,mcdermott_kangbo,sfq_genetic}. Prior work has studied the SFQ-based coherent control of qubits, and demonstrated that we can perform rotations around the $X$ or $Y$ axis by applying SFQ pulses every qubit oscillation period \cite{mcdermott_full,mcdermott_fab}. However, this approach leads to leakage to higher energy levels, thus prior work utilized a genetic algorithm to find better SFQ-based gates with low leakage and short gate time \cite{mcdermott_full,sfq_genetic}. They show that taking into account three lowest energy levels of the qubit in their model is sufficient to realize low-leakage gates using SFQ pulses.   

In \cite{mcdermott_full}, the authors envision the possibility of performing SFQ-based two-qubit gates. In \cite{alphazero}, the authors implement a quantum optimal control version of the AlphaZero learning algorithm \cite{alpha1} to optimize the quantum dynamics, and use SFQ-based optimal control as a benchmark in their study. The authors show that they can find SFQ pulse trains to do $\sqrt{ZX}$ gate with high fidelity.
However, their model of a two-qubit quantum system does not take into consideration the leakage out of the computational subspace.
Fig. \ref{fig:mot} shows the importance of taking higher energy levels into consideration when learning SFQ pulses to perform two-qubit gates. 
In each case, we report the error of the best SFQ-based two-qubit gate we find with a genetic algorithm when modeling the quantum system using $n$ energy levels.
We then simulate the learned bitstream using a model that allows for evolution to higher energy levels, and report the leakage (\cref{eq:leakage}) of the resulting gate.

We can easily find SFQ-based two-qubit gates with 0.999 fidelity with $n=2$ (consistent with prior work \cite{alphazero}). However, we find that the learned SFQ pulse train results in a gate with high leakage when allowed to evolve out of the two-level subspace. This is an expected result---prior work has shown that we need to consider $n=3$ to find low-leakage single-qubit gates \cite{sfq_genetic}. What is more surprising is that, as shown in Fig. \ref{fig:mot}, the genetic algorithm cannot find SFQ-based two-qubit gates with low leakage even with $n=5$. 
Instead, even when we learn bitstreams using $n=5$, if we simulate the same bitstreams on a system with more than $n$ energy levels, the resulting quantum evolution will leak into the additional levels, resulting in a gate with both poor accuracy and high leakage. 
Thus, unlike the single-qubit gate case, taking the higher energy level into consideration alone is not sufficient. 

In this paper, we characterize the requirements of realizing high-fidelity SFQ-based two-qubit gates. We develop quantum optimal control methods, and also investigate various qubit architectures and configurations in an attempt to engineer an SFQ-friendly quantum system that can perform high-fidelity two-qubit gates.

%% file: tex/study.tex
In this section we first discuss our methodology, followed by the results of our study on SFQ-based two-qubit gates under various qubit architectures and configurations. Then, we compare our results with that of microwave-based quantum systems.

\begin{table}[t]
\centering
\caption{The parameters used in the genetic algorithm.}
\begin{tabular}{|l|c|}
\hline
Population size & 70 \\ \hline
Selection size & 60 \\ \hline
Mutation probability & 0.001 \\ \hline
Maximum number of iterations & 200,000 \\ \hline
Target fidelity & 0.999 \\ \hline
\end{tabular}
\label{tab:gen}
\end{table}

\captionsetup[subfigure]{labelformat=empty}
\begin{figure*}[!t]
\begin{center}
\subfloat    {
\includegraphics[width=0.8\linewidth]{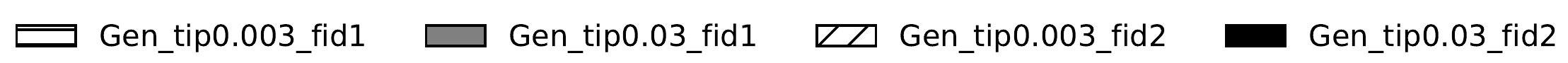}
} \\
\subfloat [(a)]    {
\includegraphics[width=0.5\linewidth]{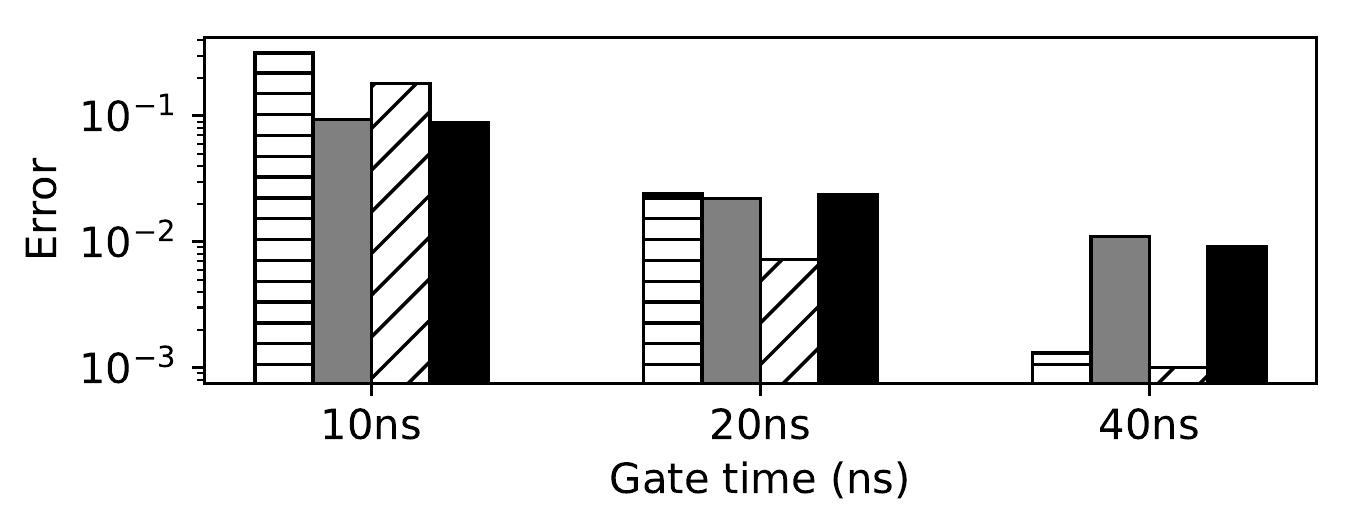}
} 
\subfloat [(b)]    {
\includegraphics[width=0.5\linewidth]{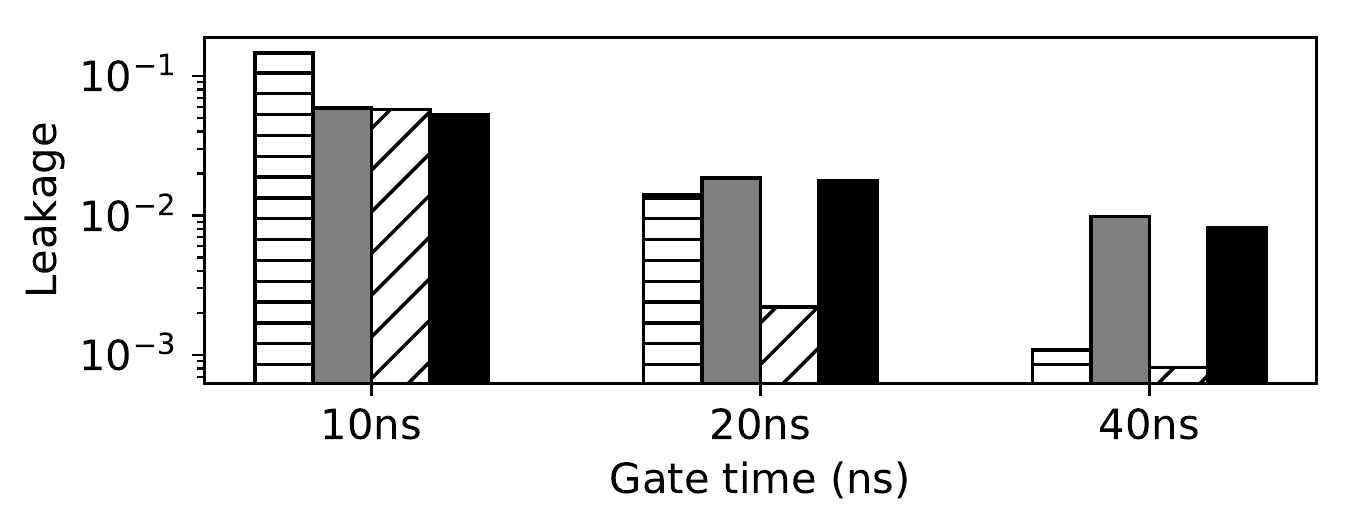}
} 
\\
\subfloat [(c)]    {
\includegraphics[width=0.5\linewidth]{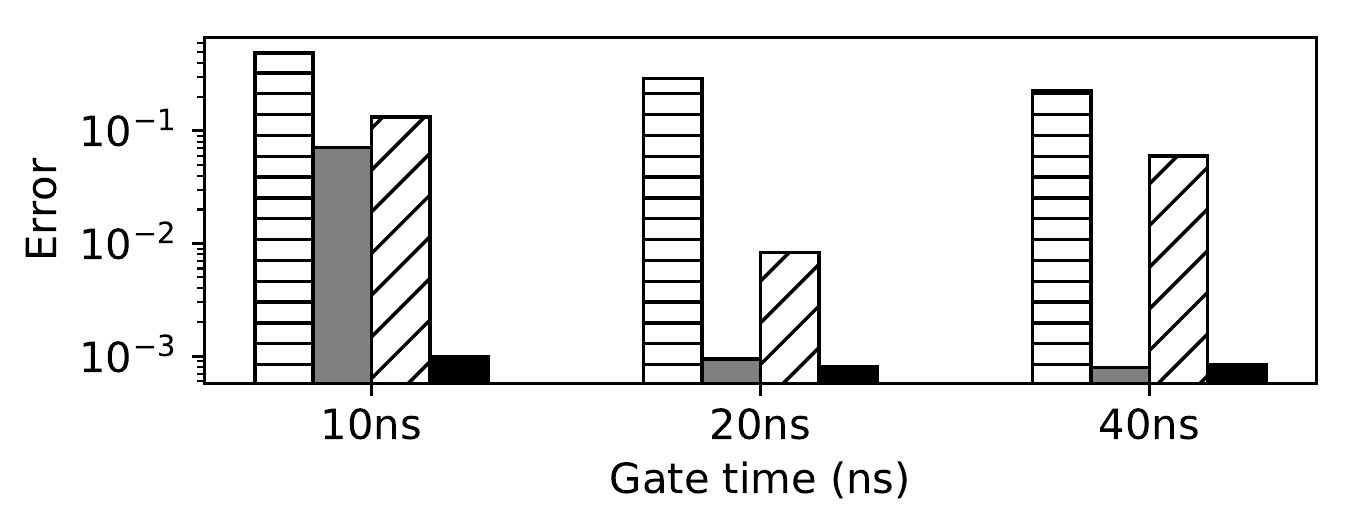}
} 
\subfloat [(d)]    {
\includegraphics[width=0.5\linewidth]{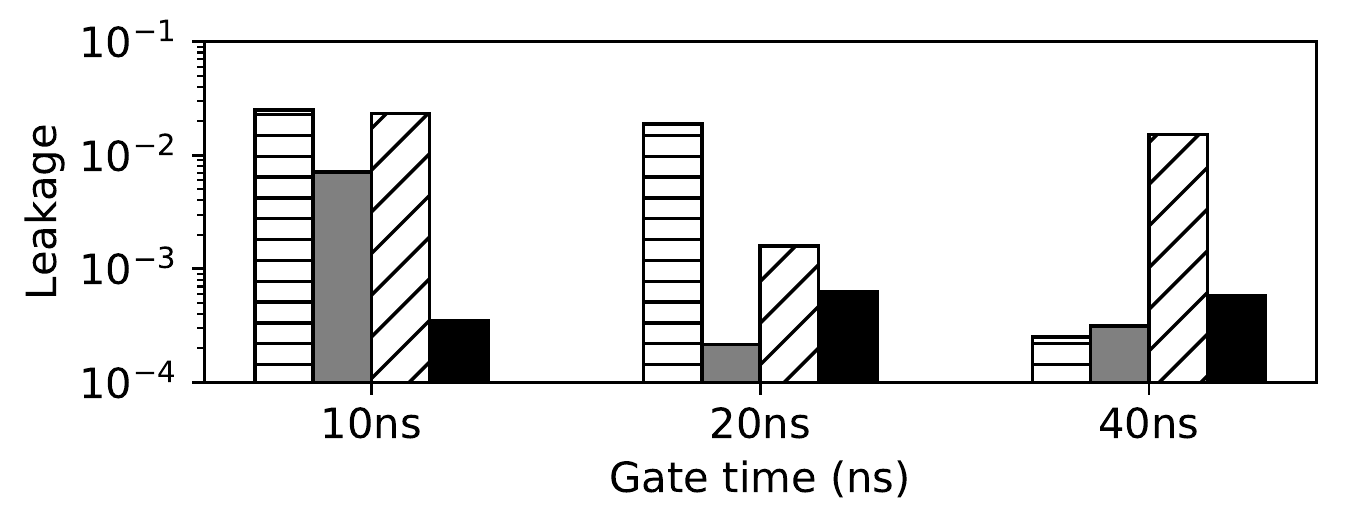}
} 
\caption{Error and leakage of the best SFQ-based CZ gate found by the genetic algorithm for transmon qubit devices with $\Omega_{x}$ control fields (plots a and b), and transmon qubit devices with $\Omega_{z}$ control fields (plots c and d). Error is calculated as $1-fidelity$, and leakage is calculated using \cref{eq:leakage}. Two different tip angles and two fidelity functions are used in our optimal control method (see Sec. \ref{sec:back} for the details of our fidelity functions). We run the simulations with $n=5$ energy levels, and suppress the population of higher energy levels in our optimal control method.}
\label{fig:transmon_cz}
\end{center}
\end{figure*}

\captionsetup[subfigure]{labelformat=empty}
\begin{figure*}[!t]
\begin{center}
\subfloat [(a)]    {
\includegraphics[width=0.96\linewidth]{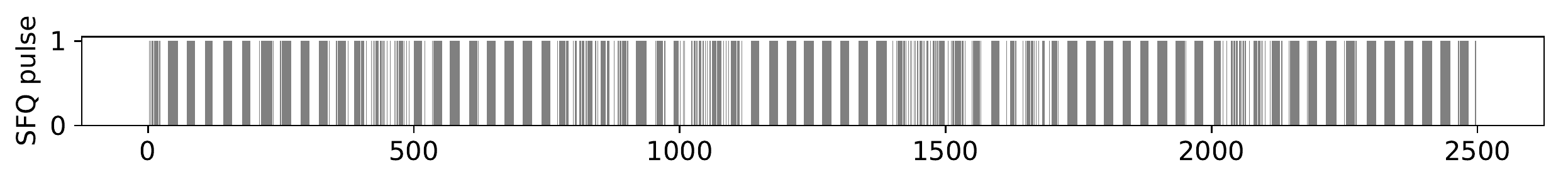}
} \\
\subfloat [(b)]    {
\includegraphics[width=0.96\linewidth]{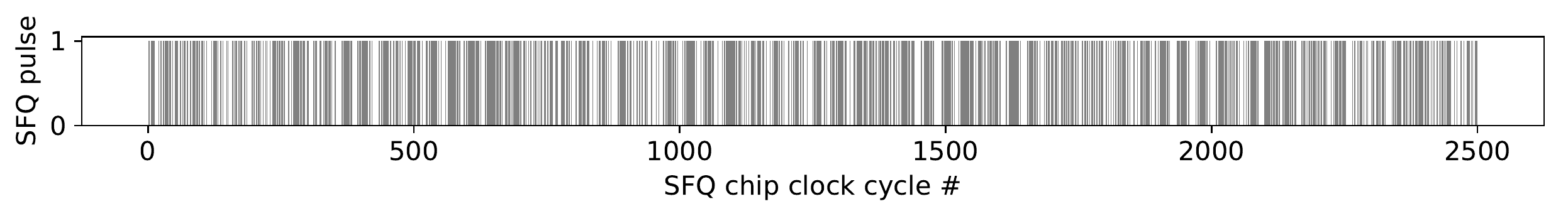}
} 
\caption{Bit representation of SFQ bitstreams applied to qubit1 (plot a) and qubit2 (plot b) on a transmon system with $\Omega_x$ control fields and 0.003 tip angle in order to realize a CZ gate with 20 ns gate time. Each SFQ chip clock cycle is 8 ps.}
\label{fig:examplebs}
\end{center}
\end{figure*}

\subsection{Methodology}
\label{sec:methodology}

We model SFQ-based quantum operations by numerically integrating the relevant system Hamiltonian (\cref{eq:bg:hsum}) over a single SFQ clock cycle for each possible combination of input pulses.
The learning algorithm then searches for pulse streams corresponding to optimal sequences of these basis operations.
In order to avoid sequences which would spill into higher levels if made available (as described in \cref{sec:motivation}), we generate each unitary evolution using extra energy levels, and then project out the extra levels after each pulse in the sequence.
The resulting non-unitarity of the evolution then gets quantified by our fidelity metrics as additional leakage, forcing the algorithm to prioritize sequences which are constrained to the given number of energy levels.

We use a variant of the genetic algorithm used in prior work \cite{alphazero} to find a train of SFQ pulses to perform quantum gates. The parameters of the genetic algorithm is summarized in Table \ref{tab:gen}. The genetic algorithm starts with a population of random SFQ pulse trains, and in each iteration, a number of parent pulse trains from the population are selected for generating new pulse trains based on a crossover function. Finally, if the fidelity is improved in the new SFQ pulse trains, they are replaced with the worst SFQ pulse trains in the population.

We use a variant of the GRAPE code used in \cite{sgrape} to find microwave pulses to perform quantum gates. We use the cost functions presented in \cite{sgrape} in order to suppress the occupation of forbidden states. Similar to the SFQ case, we set the target gate fidelity to 0.999. 

\subsection{Entangling SFQ-based two-qubit gates on transmon qubit devices}
In this section, we present the results of our analysis on transmon qubit devices. Similar to \cite{sgrape}, we use qubit frequencies of ${\omega_{01}^{(0)}}/{2\pi}=3.9$ and ${\omega_{01}^{(1)}}/{2\pi}=3.5$ GHz, anharmonicity of ${\alpha}/{2\pi}=-225$ MHz, and $n=5$ in our study on transmons. We report the results for coupling strength of ${J}/{2\pi}=50$ MHz in our main results and then perform a sensitivity analysis on the coupling strength. Note that we show the results for transmon with $\Omega_x$ control fields and transmon with $\Omega_z$ control fields separately in order to study the effectiveness of each control field on realizing entangling two-qubit gates.

\subsubsection{CZ gate on transmon qubits with $\Omega_{x}$ control fields}

Fig. \ref{fig:transmon_cz}(a) and \ref{fig:transmon_cz}(b) respectively show the error and leakage of the best SFQ pulse train found using the genetic algorithm to perform a CZ gate on transmons with $\Omega_{x}$ control fields. The leakage to higher energy levels is suppressed by the physical model employed in our optimal control method (as described in \cref{sec:methodology}). The error numbers reported in this plot take leakage to higher energy levels into consideration, thus, low error translates into low leakage as shown in Fig. \ref{fig:transmon_cz}. We run the genetic algorithm with the two fidelity functions described in \cref{sec:fidelity} (denoting \avgf/ as \avg/ and \rzf/ as \rz/), two tip angles of 0.003 and 0.03 (similar to the numbers reported in the literature \cite{mcdermott_kangbo,alphazero}), and three gate times of 10 ns, 20 ns, and 40 ns. Fig. \ref{fig:examplebs} shows an example of the SFQ bitstreams that are learned using the genetic algorithm.

Our results show that it is hard to find high-fidelity CZ gates while suppressing the leakage to higher energy levels using the 0.03 tip angle with either fidelity function.
By decreasing the tip angle to 0.003, we are able to realize a CZ gate with 0.999 fidelity and 40 ns gate time. Decreasing the tip angle means the amount of energy deposited into the qubit with each SFQ pulse decreases, thus, the required gate time to perform high-fidelity quantum operations increases.  

In general, \rz/ results in better SFQ-based pulse trains than \avg/, indicating that the broader target provided by \rz/ is indeed more friendly to the highly-constrained nature of SFQ-based gate implementation.

\captionsetup[subfigure]{labelformat=empty}
\begin{figure}[!t]
\begin{center}
\subfloat    {
\subfloat []    {
\includegraphics[width=0.95\linewidth]{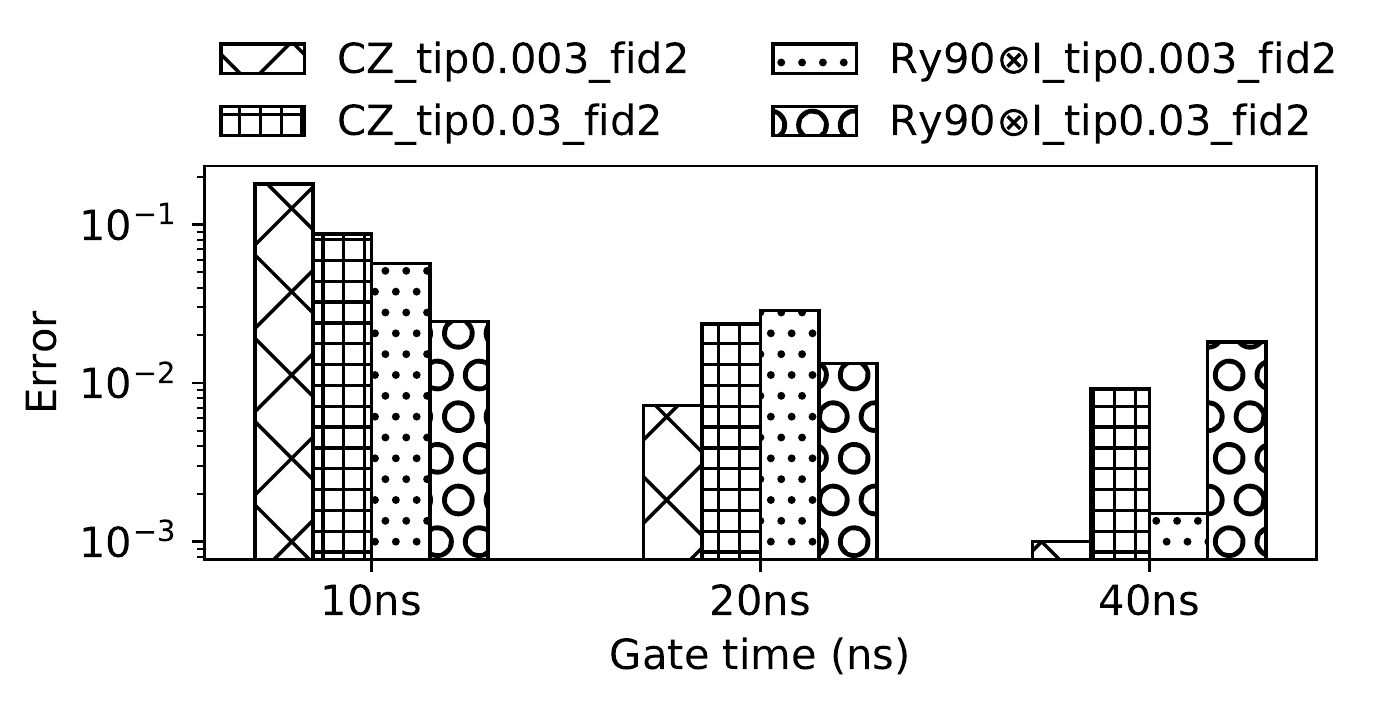}
} }
\caption{Error of the best SFQ-based CZ gate (entangling two-qubit gate) and Ry90$\otimes$I  gate (non-entangling two-qubit gate) found by the genetic algorithm for transmon qubit devices with only $\Omega_{x}$ control fields.}
\label{fig:full}
\end{center}
\end{figure}

\subsubsection{CZ gate on transmon qubits with $\Omega_{z}$ control fields}

Fig. \ref{fig:transmon_cz}(c) and \ref{fig:transmon_cz}(d) show the error and leakage results of transmon devices with $\Omega_z$ control fields, respectively. Here, we apply a $\Omega_z$ control field only to qubit2 (which is sufficient to realize high-fidelity CZ gates). We observe a significant reduction in the amount of leakage to higher energy levels in the case of transmon devices with $\Omega_z$ control fields compared to that of transmon devices with $\Omega_x$ control fields. Since the leakage is low in this case, we can afford to use higher tip angles in order to perform fast gates. Our results show that we can realize high-fidelity CZ gates with $<$0.001 error and $<$0.001 leakage with 0.03 tip angle and 10 ns gate time with \rz/ (longer gate time is required with \avg/). 

Our findings show that $\Omega_z$ control field with 0.003 tip angle is not sufficient to realize high-fidelity CZ gates. 
However, the gates that we find do have low leakage in some cases;
although low error translates to low leakage because we take into consideration the leakage to higher energy levels in calculating the error values, the opposite is not necessarily true (for example, the identity gate has high error if we calculate its overlap with CZ gate, but it has low leakage to higher energy levels).

\captionsetup[subfigure]{labelformat=empty}
\begin{figure}[!t]
\begin{center}
\subfloat [(a)]    {
\includegraphics[width=0.96\linewidth]{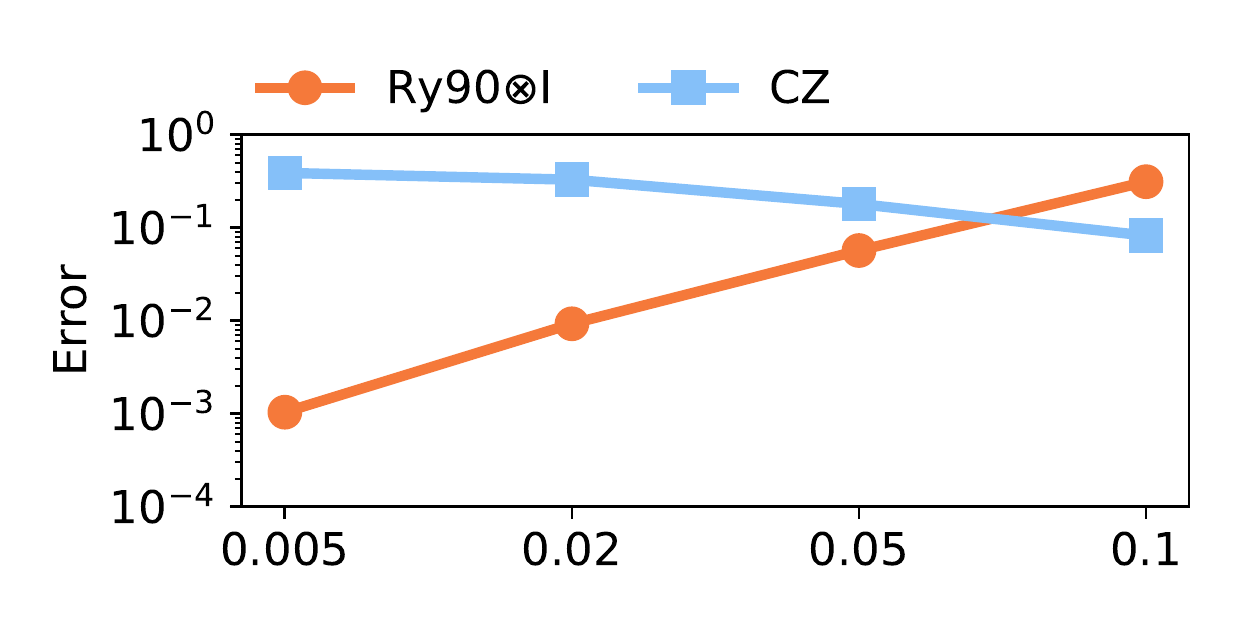}
} \\
\subfloat [(b)]    {
\includegraphics[width=0.96\linewidth]{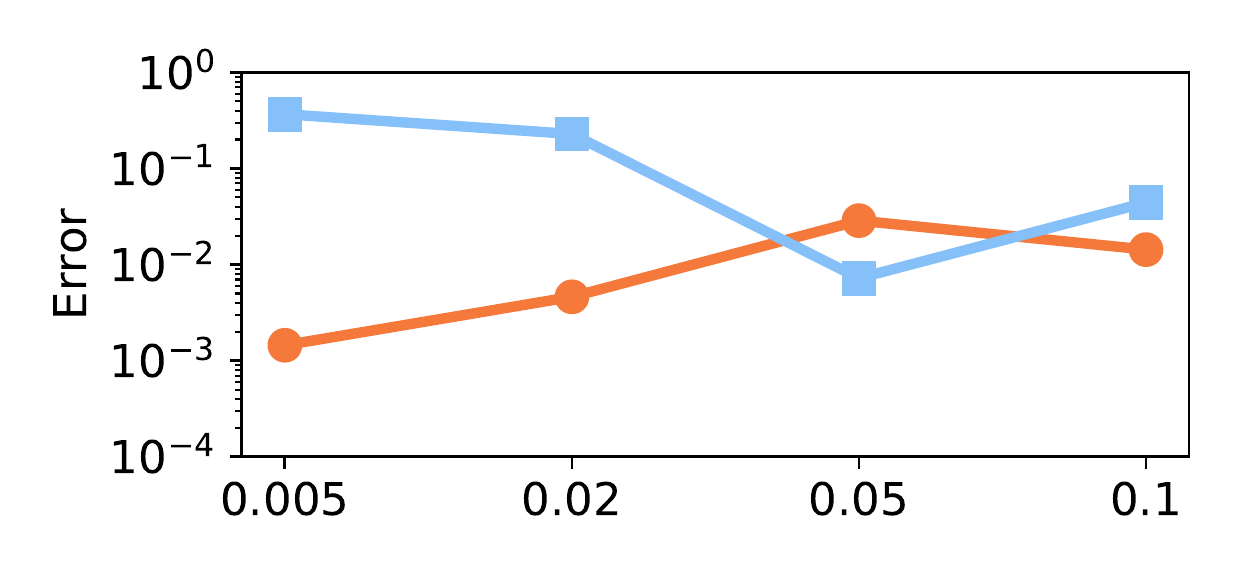}
} 
\\
\subfloat [(c)]    {
\includegraphics[width=0.96\linewidth]{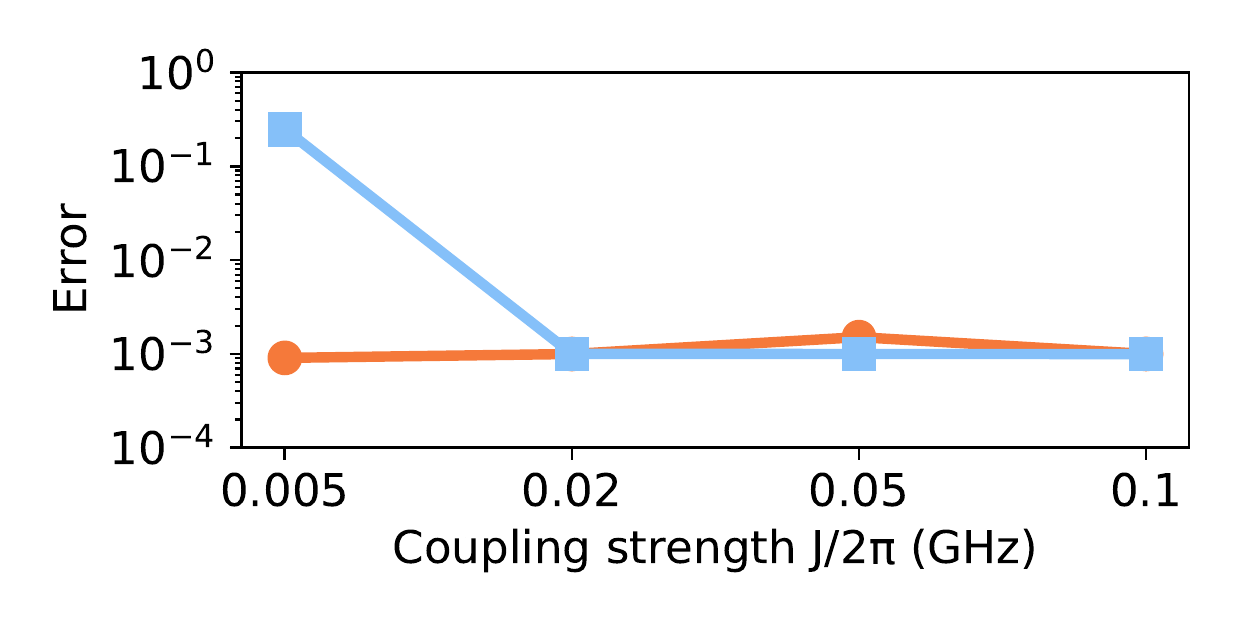}
} 
\caption{Sensitivity analysis on qubit coupling strength in transmon system with $\Omega_x$ control fields. The results are shown for 0.003 tip angle and 10 ns (plot a), 20 ns (plot b), and 40 ns (plot c) gate times. The SFQ bitstreams are learned with \rz/.}
\label{fig:jsense}
\end{center}
\end{figure}

\subsection{Realizing both entangling and non-entangling SFQ-based two-qubit gates on transmon devices}
So far, we demonstrated that we can realize high-fidelity CZ gates with low leakage and short gate time using transmon qubit devices, which is a promising result. However, it is essential to ensure that we can also realize high-fidelity one-qubit gates in our two-qubit quantum system (i.e., non-entangling two-qubit gates). Next, we study the requirements of a system that can perform both entangling and non-entangling SFQ-based two-qubit gates. Transmon system with only $\Omega_z$ control fields is suitable to realize high-fidelity entangling two-qubit gates, but it does not provide enough control to perform arbitrary single-qubit gates, which as expected leads to non-entangling two-qubit gates with high error ($>10^{-1}$ error). Thus, we need more than just $\Omega_z$ control fields to realize both entangling and non-entangling two-qubit gates. Next, we investigate two systems as possible candidates to achieve this goal.

\captionsetup[subfigure]{labelformat=empty}
\begin{figure*}[!ht]
\begin{center}
\subfloat    {
\includegraphics[width=0.8\linewidth]{fig/legend.pdf}
} \\
\subfloat [(a)]    {
\includegraphics[width=0.5\linewidth]{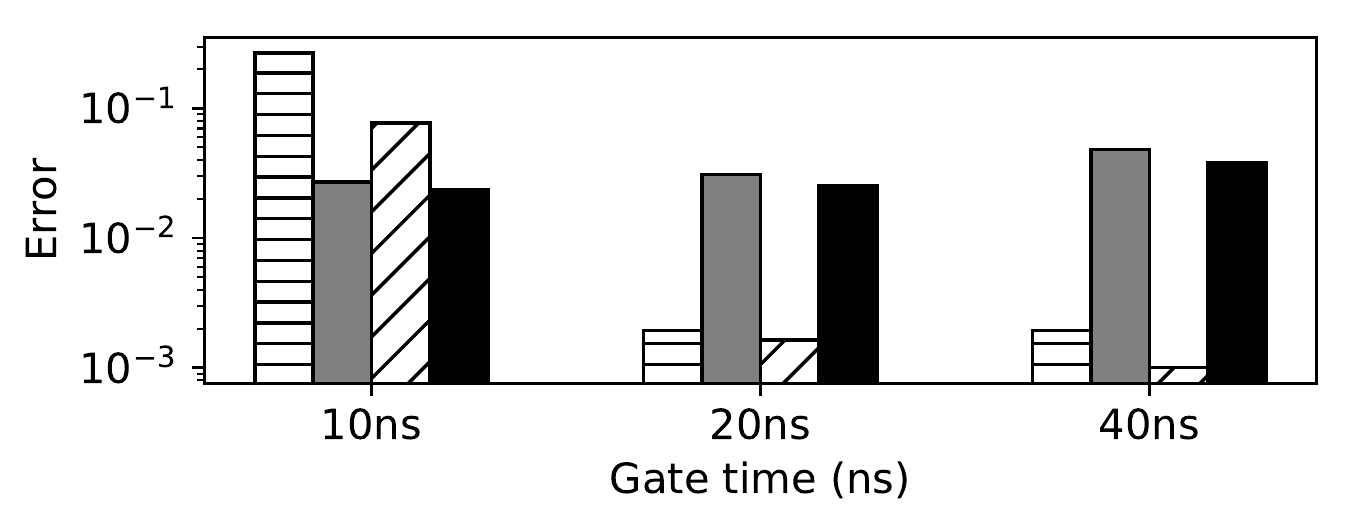}
} 
\subfloat [(b)]    {
\includegraphics[width=0.5\linewidth]{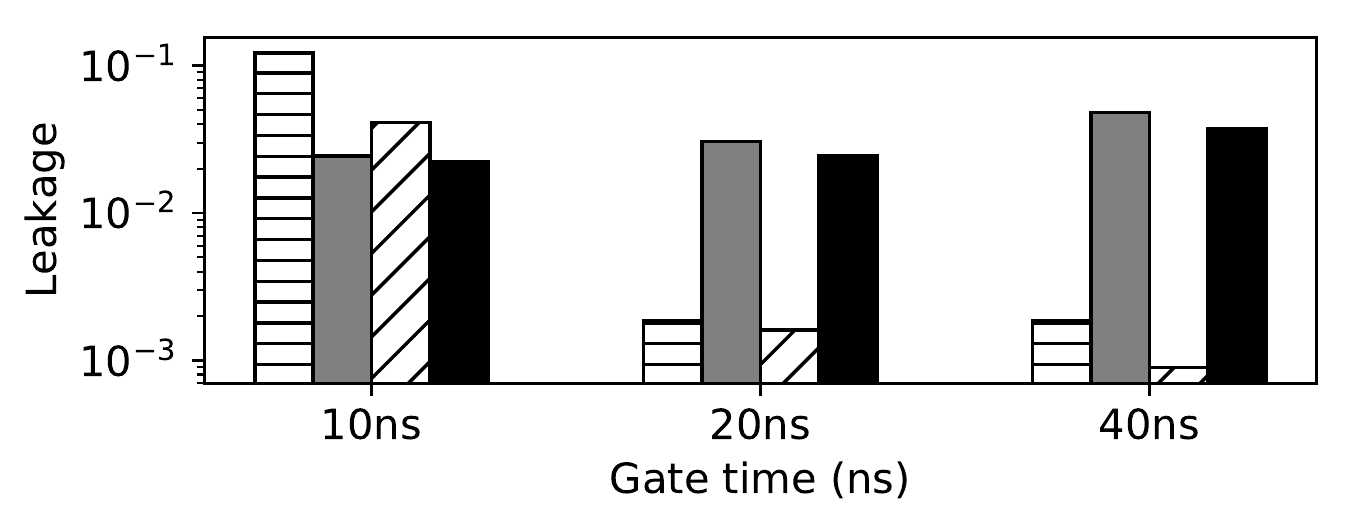}
} 
\\
\subfloat [(c)]    {
\includegraphics[width=0.5\linewidth]{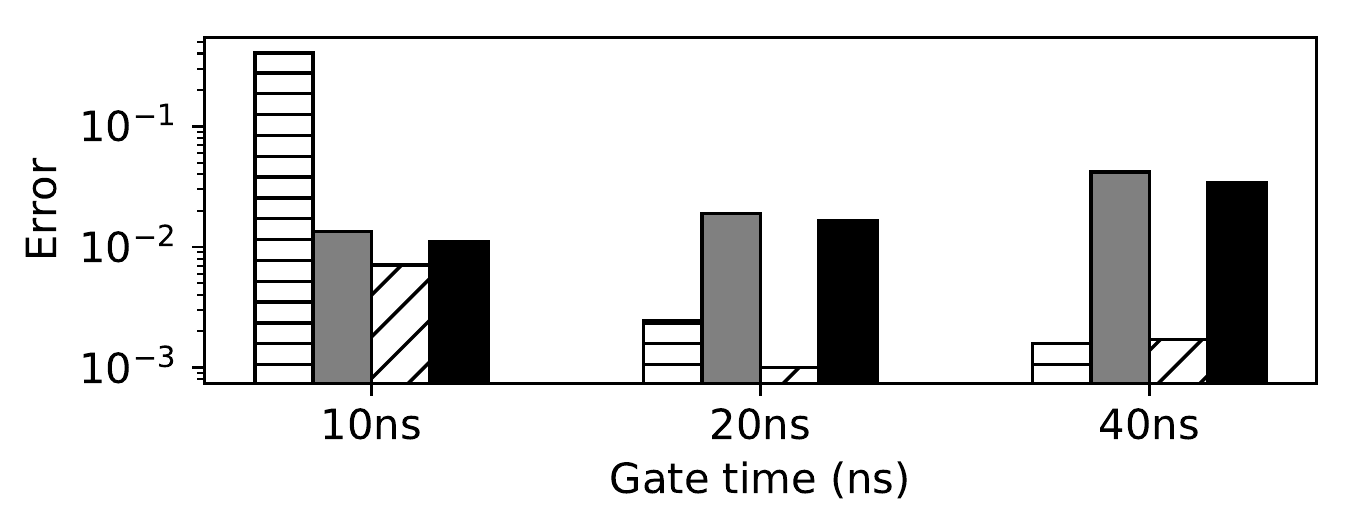}
} 
\subfloat [(d)]    {
\includegraphics[width=0.5\linewidth]{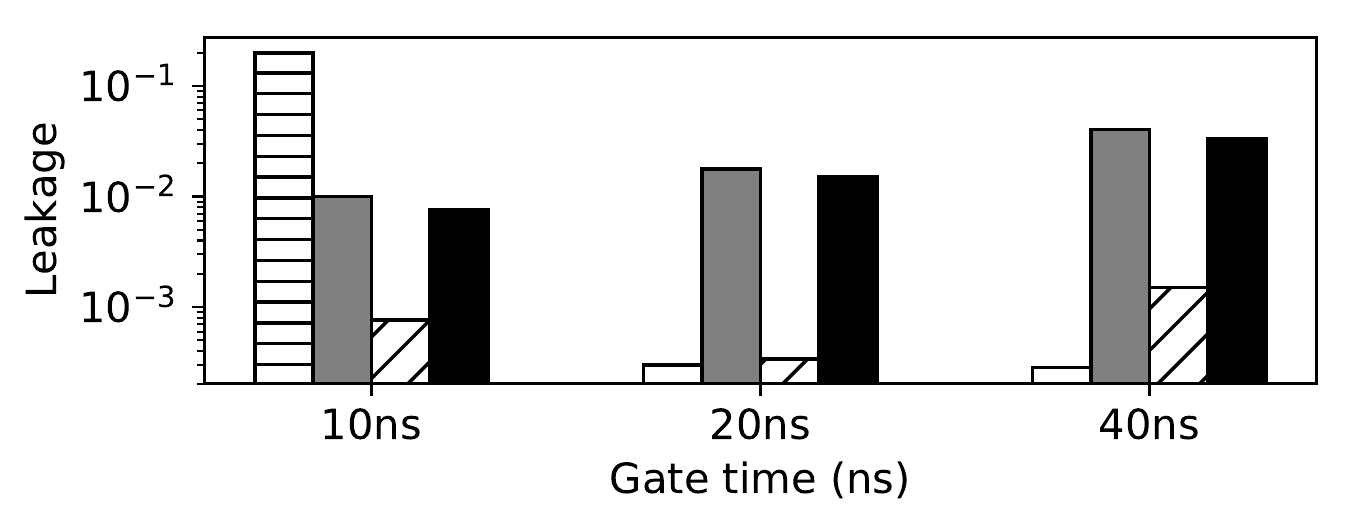}
} 
\caption{Error and leakage of the best SFQ-based CZ gate  (plots a and b) and Ry90$\otimes$I gate  (plots c and d) found by the genetic algorithm for fluxonium qubit devices with $\Omega_{x}$ control fields.}
\label{fig:flux}
\end{center}
\end{figure*}

\subsubsection{Transmon with $\Omega_x$ control fields}
Prior work demonstrated SFQ-based single-qubit gates with $<$20 ns gate time on transmon devices with only $\Omega_x$ control fields \cite{mcdermott_kangbo,sfq_genetic}. In addition, we showed earlier that we can use transmon devices with $\Omega_x$ control fields to perform high-fidelity CZ gates with 40 ns gate time and low tip angle. A natural question arises: can we engineer a transmon system with $\Omega_x$ control fields that can perform both entangling and non-entangling SFQ-based two-qubit gates with high fidelity? Fig. \ref{fig:full} shows the error results of the best SFQ-based CZ gate (entangling two-qubit gate) and Ry90$\otimes$I gate (non-entangling two-qubit gate) found on transmon system with $\Omega_x$ control fields using the genetic algorithm. Our results show that we can realize both CZ and Ry90$\otimes$I gates with high fidelity with 0.003 tip angle and 40 ns gate time. 

One interesting observation is that the gate time of Ry90$\otimes$I is longer than the gate time of the SFQ-based single qubit gates reported in prior work \cite{mcdermott_kangbo,sfq_genetic}. In general, it is more challenging to realize precise single-qubit gates in a two-qubit system compared to the one-qubit systems because of crosstalk with the neighbor qubit. We can reduce crosstalk and achieve faster single-qubit gate times by reducing the coupling strength ($J$), however this will in turn complicate the realization of two-qubit entangling gates. Fig. \ref{fig:jsense} shows a sensitivity analysis on the coupling strength. Our results show that realizing high-fidelity CZ gate requires higher coupling strengths and realizing high-fidelity Ry90$\otimes$I gate requires lower coupling strengths. A coupling strength of $J/2\pi=50$ MHz is a sweet spot that works well for both CZ and Ry90$\otimes$I gates in our results.

\subsubsection{Transmon with both $\Omega_x$ and $\Omega_z$ control fields}
One possible configuration is to dedicate both $\Omega_x$ and $\Omega_z$ control fields to the transmon qubit devices, which would potentially lead to SFQ-based gates with higher fidelity and shorter gate time.
However, we note that this comes at the cost of hardware complexity and heightened sensitivity to magnetic flux noise.

\begin{figure*}[!ht]
\begin{center}
\includegraphics[width=0.6\linewidth]{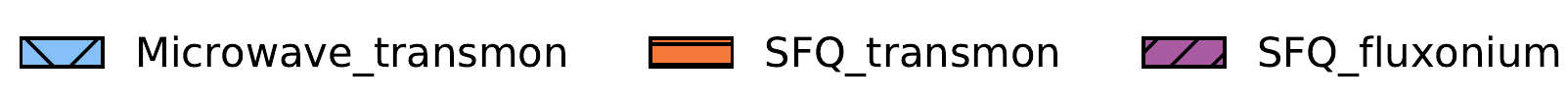} \\
\subfloat   [(a)] {
\includegraphics[width=0.5\linewidth]{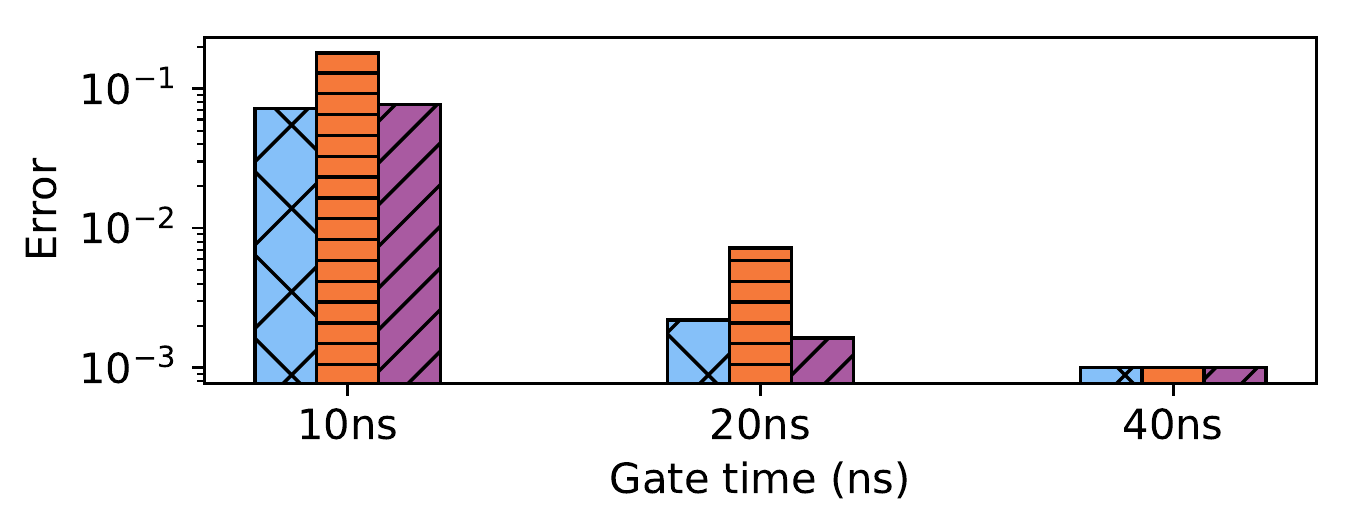}
}
\subfloat [(b)] {\includegraphics[width=0.5\linewidth]{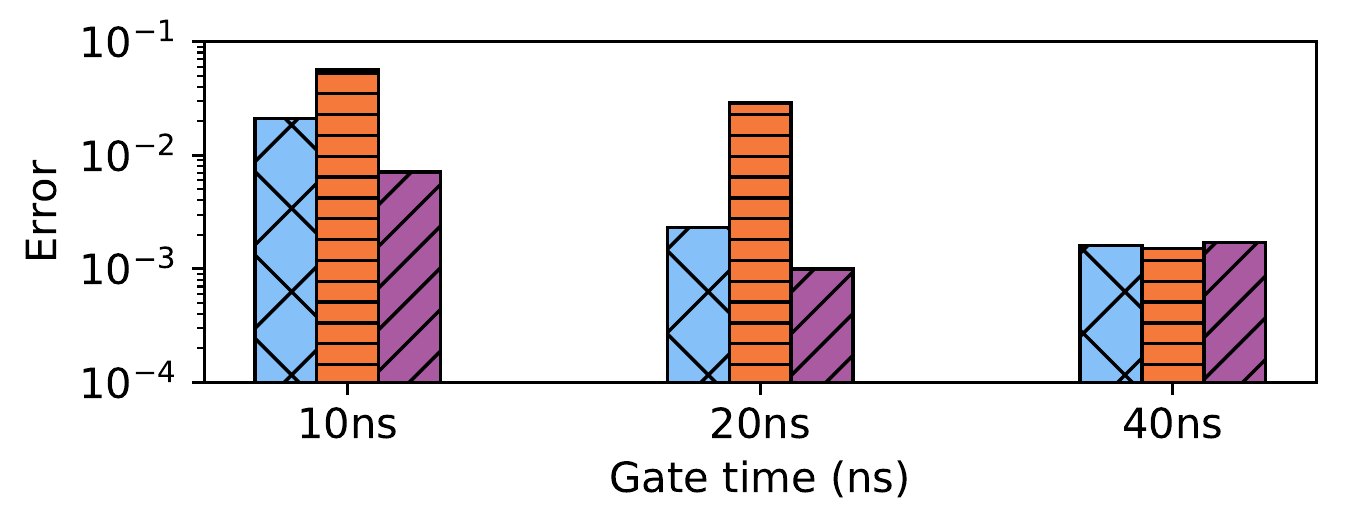}}
\caption{Error comparison between microwave-based gates obtained using Grape code and SFQ-based gates obtained using genetic algorithm (with 0.003 tip angle). The results are reported for CZ gate (plot a) and Ry90$\otimes$I gate (plot b).}
\label{fig:microwave}
\end{center}
\end{figure*}

\subsection{SFQ-based two-qubit gates on fluxonium qubit devices}
In this section, we investigate fluxonium qubit devices as a possible candidate to realize both SFQ-based entangling and non-entangling gates with low leakage and short gate time using $\Omega_{x}$ control fields.
Our model for the fluxonium devices assumes qubit1 (qubit2) is configured with $\ej=5.5$ (5.7), $\ec=1.5$ (1.2) and $\el=1.0$, and 
a static $\bfield=\pi$ external flux.
Fig. \ref{fig:flux} shows the results of our study on fluxonium devices with $\Omega_{x}$ control fields.

Fig. \ref{fig:flux}(a) and \ref{fig:flux}(b) show the error and leakage results of an SFQ-based CZ gate, respectively. Our results show that we can realize high-fidelity CZ gates with a gate time of 20 ns thanks to the low leakage of fluxonium devices. Similar to the case of transmons with $\Omega_{x}$ control fields, better results are achieved with lower tip angle.
Fig. \ref{fig:flux}(c) and \ref{fig:flux}(d) show the error and leakage results of an SFQ-based Ry90$\otimes$I gate, respectively. Our results show that the genetic algorithm can find high-fidelity gates with 20 ns gate time.

The fluxonium results show the feasibility and effectiveness of both entangling and non-entangling two-qubit gates with short gate time and low error and leakage using only $\Omega_{x}$ control fields.

\subsection{Comparison with microwave-based gates}

Finally, we compare our results with that of microwave-based gates obtained from the GRAPE algorithm \cite{sgrape}. Fig. \ref{fig:microwave} shows the error results for CZ gate and Ry90$\otimes$I gate for three designs: (1) microwave-based design with transmon devices; (2) SFQ-based design with transmon devices; (3) SFQ-based design with fluxonium devices.
We learn the SFQ pulse trains with the \rzf/ function. In the microwave case, \avgf/ is sufficient to realize high-fidelity gates.

The results reported in Fig. \ref{fig:microwave} show that the SFQ-based design with fluxonium has similar error to that of the microwave-based design. The SFQ-based design with transmons has similar error to the other two systems for 40 ns gate time, and higher error than the other two systems for 10 ns and 20 ns gate times.
The comparison results show that we can perform high-fidelity SFQ-based gates with similar gate time and gate fidelity to that of microwave-based system. Thus, SFQ is a promising approach to implement classical controllers as they can deliver quantum computers with both high scalability and high fidelity.

%% file: tex/conclusion.tex
Superconducting Single Flux Quantum (SFQ) is a classical logic technology which is proposed in the literature to implement in-fridge classical controllers in order to maximize the scalability of quantum computers. In this paper, we demonstrate the first thorough analysis of SFQ-based two-qubit gates -- a key remaining step in realizing SFQ-based universal quantum computing. Our results show that despite the severe challenges of realizing SFQ-based two-qubit gates, they are both feasible and effective if we carefully design our quantum optimal control method and qubit architecture. We characterize the requirements of such gates, and carefully engineer SFQ-friendly quantum systems that can perform both two-qubit gates and single-qubit gates with high fidelity on a system with fixed coupling (tunable coupling would potentially provide further isolation and less crosstalk, however, further research is required to investigate the effectiveness of such couplers on an SFQ-based system). More importantly, we demonstrate that the fidelity and gate time of these gates are comparable to that of microwave-based gates -- these results show that SFQ approach can potentially not only increase the scalability of quantum machines but also maintain the fidelity and effectiveness of quantum gates, thus SFQ is a promising approach to implement classical controllers for scalable quantum machines.